\documentclass[sigconf,screen]{acmart}
\renewcommand\footnotetextcopyrightpermission[1]{} 
\AtBeginDocument{%
  \providecommand\BibTeX{{%
    Bib\TeX}}}

\usepackage{algorithmic}
\usepackage{graphicx}
\usepackage{textcomp}
\usepackage{xcolor}
\usepackage{tabularx}
\def\BibTeX{{\rm B\kern-.05em{\sc i\kern-.025em b}\kern-.08em
    T\kern-.1667em\lower.7ex\hbox{E}\kern-.125emX}}
\usepackage{tcolorbox}
\usepackage{multirow}
\usepackage{enumitem}
\setlist[itemize,1]{left=0pt} %

\begin{document}

\newcommand{\realsearchgoal}{\textbf{\textit{ The goal of this study is to assist practitioners and policymakers in making informed decisions on which \practices to adopt by evaluating the relationship between software \practices adoption and security outcome metrics.
}}}

\newcommand{\rqPatternicse} {\textbf{RQ1}: Is higher adoption of aggregated security practices associated with measurable improvements in security outcomes? }

\newcommand{\HOne} {\textbf{H1}: Packages that exhibit a good security posture through the adoption of security practices will demonstrate improved security outcomes.}%

\newcommand{\rqOutcomeicse}{\textbf{RQ2}: Which security practices are most strongly associated with security outcomes? }

\newcommand{\HTwo}{\textbf{H2}: The strength of associations between individual security practices and security outcomes varies, indicating that certain practices have a greater influence than others.}

\newcommand{\RQThree}{\textbf{RQ3}: Which security practices have the strongest measurable impact on security outcomes? }

\newcommand{\vulcount}{Vul\_Count}
\newcommand{\vulcountspace}{Vul\_Count }

\newcommand{\practice}{security practice }

\newcommand{\practices}{security practices }
\newcommand{\totalpackage}{145,817}
\newcommand{\totalpackagespace}{145,817 }
\newcommand{\totalMTTR}{22,412}
\newcommand{\totalMTTRspace}{22,412 }

\title{Assumptions to Evidence: Evaluating Security Practices Adoption and Their Impact on Outcomes in the npm Ecosystem}

\author{Nusrat Zahan}
\email{nzahan@ncsu.edu}
\affiliation{%
  \institution{North Carolina State University}
  \city{Raleigh}
  \state{NC}
  \country{USA}
}
\author{Imranur Rahman}
\email{irahman@ncsu.edu}
\affiliation{%
  \institution{North Carolina State University}
  \city{Raleigh}
  \state{NC}
  \country{USA}
}
\author{Laurie Williams}
\email{lawilli3@ncsu.edu}
\affiliation{%
  \institution{North Carolina State University}
  \city{Raleigh}
  \state{NC}
  \country{USA}
}

\renewcommand{\shortauthors}{Zahan et al.}

\begin{abstract}
Practitioners often struggle with the overwhelming number of security practices outlined in cybersecurity frameworks for risk mitigation. Given the limited budget, time, and resources, practitioners want to prioritize the adoption of security practices based on empirical evidence.
\textit{\realsearchgoal} To do this, we analyzed the adoption of security practices and their impact on security outcome metrics across 145K npm packages. We selected the OpenSSF Scorecard metrics to automatically measure the adoption of security practices in npm GitHub repositories. We also investigated project-level security outcome metrics: the number of open vulnerabilities~(\vulcount), mean time to remediate~(MTTR) vulnerabilities in dependencies, and mean time to update~(MTTU) dependencies. We conducted regression and causal analysis using 11 Scorecard metrics and the aggregated Scorecard score (computed by aggregating individual security practice scores) as predictors and \vulcount, MTTR, and MTTU as target variables. 
Our findings reveal that aggregated adoption of security practices is associated with 5.2 fewer vulnerabilities, 216.8 days faster MTTR, and 52.3 days faster MTTU. Repository characteristics have an impact on security practice effectiveness: repositories with high security practices adoption, especially those that are mature, actively maintained, large in size, have many contributors, few dependencies, and high download volumes, tend to exhibit better outcomes compared to smaller or inactive repositories. Among individual security practices, Code Review, Contributors from diverse organizations, License, CI-Tests, and Pinned Dependencies show strong
associations with security outcomes.

\end{abstract}

\maketitle

\section{Introduction}
The growing reliance on third-party dependencies in modern software applications expands the attack surface, exposing applications to CVE (Common Vulnerabilities and Exposures catalog) exploits or malicious code injection, facilitating software supply chain (SSC) attacks~\cite{williams2024research}. For example, CVEs have grown by 463\% in the last 10 years, and 704,102 malicious packages have been proactively identified since 2019~\cite{sonatype2024}. Practitioners are falling behind in managing vulnerabilities in dependencies due to growing dependency complexity and resource constraints, with mean time to remediate~(MTTR) vulnerabilities increasing from under 25 days in 2017 to over 400 days by 2024~\cite{sonatype2024}.

SSC security standards and frameworks (e.g. ~\cite{EO_2021, ENISA_CRA_2024, EO2025Cybersecurity, souppaya2022secure, Microsoft_framework, Williams2024PSSCRM}) have been proposed by industry and government agencies to help organizations adopt security practices that mitigate SSC security risks.  However, historical experience demonstrates the risks of adopting security practices without empirical validation. For example, long password complexity requirements, such as mandating uppercase letters, numbers, and symbols, were long assumed to enhance security~\cite{adams1999users, morris1979password}. However, empirical studies~\cite{dekoven2022measuring, dekoven2019measuring, lain2024content} later revealed that such policies often led users to create predictable passwords by encouraging reuse and minor modifications. Similar theoretical assumptions may persist in SSC security, where different practices are widely promoted without comprehensive empirical validation. Security success is often measured through adoption metrics—e.g., the number of teams that adopted code review or the number of vulnerabilities detected or fixed, rather than actual security outcomes. Therefore, little is known about the effectiveness of best practices in improving security. The absence of empirical evidence arises from multiple systemic challenges: widespread disagreement about best practices among experts~\cite{reeder2017152,redmiles2020comprehensive}, poor evidence even where consensus exists~\cite {shostack2019empirical, thompson2017large}, and dependence on intuition and convention rather than empirical evidence.  

The cybersecurity community's reliance on perceived common sense has created an information-poor environment characterized by conflicting guidance and expert disagreement about fundamental priorities~\cite{reeder2017152,redmiles2020comprehensive, national2025cyber}. Meanwhile, organizations have limited budgets, time, and resources~\cite{sammak2023developers, hamer2025closing}, yet practitioners currently lack the empirical evidence needed to prioritize security practices that mitigate security risk. Therefore, developing empirical research to evaluate the security outcome of recommended practices is both a research priority and a fundamental prerequisite for evidence-based security decision-making~\cite{scott2025counting, national2025cyber}. \textit{\realsearchgoal} %
We address the following research questions (RQs):
\begin{itemize}
    \item \rqPatternicse %
    \item \rqOutcomeicse %
\end{itemize}
To that end, the study investigated the relationship between publicly available data on security practice adoption and security outcomes metrics. For large-scale analysis, we focus on security practice adoption that can be measured automatically. 
We selected the OpenSSF Scorecard tool~\cite{Scorecard} that automatically collects 18 metrics on the adoption of security practices in GitHub repositories. The Scorecard metrics are valuable for collecting ecosystem-scale data, as well as for evaluating the security posture of project dependencies. Examples of security practices that can be measured by Scorecard include maintainers conducting code reviews before merging pull requests, using a static analyzer tool, and others.  For each metric, the Scorecard tool assigns a score of 0-10 based on pre-defined criteria. The tool also computes an \textit{aggregate security score} as a risk-weighted average of the individual security practice metrics, weighted by risk. For security outcomes, we selected project-based outcome metrics, where a project is defined as a software package that includes all components in its dependency graph. Specifically, the number of open vulnerabilities (\vulcount), the mean time to remediate (MTTR) vulnerable and outdated dependencies, and the mean time to update (MTTU) outdated dependencies were used. %

To address RQ1, we investigate the aggregate adoption of security practices and their impact on security outcomes. We developed a regression model to assess the relationship using Scorecard \textit{aggregate security score} and security outcome metrics. Additionally, we applied Propensity Score Matching (PSM), a statistical matching technique used in observational studies, to estimate causal effects and mitigate the potential confounding effects based on repository characteristics. For RQ2, we employed a regression model and feature importance-ranked techniques to identify which individual security practices have the strongest associations with security outcomes. To further validate these findings, we used PSM to estimate the effect of individual security practices on security outcomes, isolating their impact while accounting for differences in repository characteristics. The study has the following contributions:

\begin{itemize}
\item Introduction of project-based security outcome metrics.
\item Application of industry-standard metrics, MTTR and MTTU, to assess security outcomes.
\item A proposed model that can be used to explore the relationship between security practice and security outcome metrics.
\item Empirical evidence on how security practices impact the security outcome of npm packages.
\end{itemize}

The paper is organized as follows: Section \ref{def} contains the definition of security practice and security outcome; Section~\ref{design} discusses the design of our study; Section \ref{metrics} discusses the metrics used in our study; and Section~\ref{dataset-contruction} discusses our dataset construction steps. Section \ref{sc:data-preporocess}, Section~\ref{sc:Rq1} and Section~\ref{sc:Rq2} describe the data processing, methodology, and results. We close with a discussion of our study (Section~\ref{sc:disc}) and related work in Section \ref{sc:Related work}.

\section{Definition} \label{def}
\textbf{Security practices} are actions, procedures, and technical controls that an organization can adopt with the goal of preventing, detecting, and responding to threats against information systems. Security practices encompass both technical and organizational measures that are typically standardized and carried out by individuals or teams as part of an organization’s efforts to protect its information, resources, and operations~\cite{king2000best}. %

\textbf{Security outcomes} are quantifiable indicators that reflect the effectiveness of security practices implemented within an information system~\cite{EO2025Cybersecurity}. Outcome metrics assess whether implemented security practices are achieving their intended security risk management objectives, specifically reducing harm to information systems and guiding continuous security improvement. Security outcome metrics may include indicators such as reductions in attack frequency, security incidents, or financial losses; decreased vulnerability exposure (e.g., lower vulnerability counts), and faster response times, which shorten the window of exposure to threats.

\section{Experiment design} \label{design}
In this section, we outline the experimental design and dataset used to investigate our research questions. To answer RQ1 and RQ2, we define Scorecard security practices metrics and aggregate security score as predictor variables and security outcome metrics as target variables. We also incorporate control variables to account for potential confounding project characteristics. We leveraged machine learning and causal analysis techniques to examine the relationship between practices and outcomes. Our analysis is based on data collected from npm packages and their corresponding GitHub repositories, using dependency graphs to define project scope. The dataset construction integrates multiple data sources and applies systematic inclusion and exclusion criteria to ensure reliability and reproducibility. Section~\ref{metrics} details the selection of metrics used in our study, while Section~\ref{dataset-contruction} describes how we collected the dataset.

\subsection{Metrics Utilized in the Study} \label{metrics}
In this section, we discuss our metrics selection process, including security practices, security outcomes, and control variables. 

\subsubsection{\textbf{Security Practice Metrics}}  
To assess the adoption of security practices, the OpenSSF Scorecard tool~\cite{Scorecard} was employed. The tool automatically evaluates a set of security practices implemented in GitHub repositories, based on observable project metadata and activity. %
Table ~\ref{tab:scorecards} contains the list of 18 security practices measured by Scorecard. The tool also computes an \textit{aggregate security score} to assess the overall security of a repository, computed as a risk-weighted average of individual security practice metrics.  Scorecard was selected due to its widespread adoption and recognition in both academic and industry contexts. In academic research, Scorecard has been the subject of empirical evaluations, including SSC risk measurement~\cite{zahan2023software, mounesan2023exploring, butler2024links, salvievaluating, siddiqui2024elevating, kancharoendee2025categorizing, wu2025libvulnwatch,rydin2025evaluating}, surveys on practitioners' perspectives on Scorecard~\cite {elder2024applying,rydin2025evaluating}, and qualitative studies~\cite{zahan2023openssf} examining its applicability and effectiveness. In industry, Scorecard has been adopted as part of standard security frameworks and requirements~\cite{Williams2024PSSCRM, openssf_osps_baseline_2025} and has also been used in reports by organizations, such as Sonatype and Veracode, to assess the state of open-source software ecosystems~\cite{sonatype_2022, eng2024quantifying}. Adoption by major projects such as Kubernetes, Node.js, and Apache, along with integration into platforms like Ortelius~\cite{ortelius2024dashboard} and Deps.dev~\cite{OSI}, also highlights wide adoption. %

\subsubsection{\textbf{Security Outcome Metrics}}  
A comprehensive understanding of security outcome metrics remains an open research challenge. Although defining and validating such metrics is inherently challenging, our work represents a step toward advancing that goal. In this study, we selected project-level outcome metrics because prior studies~\cite{zahan2023software} showed that only a small fraction of packages (0.1\% of the npm ecosystem) have reported vulnerabilities at the individual package level. Limited availability of vulnerability data prevents models from learning meaningful relationships between security practices and security outcomes. Therefore, we leveraged three project-level outcome metrics, where a project is defined as an individual package and its dependencies. The security risk of a project decreases when a vulnerable component is replaced with a fixed one, providing a measurable indication of improved security outcomes, reflected in fewer open vulnerabilities and reduced exposure windows due to faster response times.

\begin{itemize}
\item \textbf{\vulcount}: A measure quantifying the number of publicly disclosed or reported open security vulnerabilities within a package and its full dependency tree, including both direct and transitive dependencies.

\item \textbf{MTTR}: A measure of the average aggregated time a package uses outdated and vulnerable direct dependencies in its lifetime~\cite{rahman2024characterizing}. MTTR quantifies how long vulnerable dependencies remain outdated after a new version with a security fix has been released.
\item \textbf{MTTU}: A measure of the average aggregated time a package uses outdated direct dependencies in its lifetime~\cite{rahman2024characterizing}. MTTU quantifies how long dependencies remain outdated after a new version has been released. 
\end{itemize}

The number of vulnerabilities is widely used in academic research, industry, and policy as outcome metrics~\cite{EO2025Cybersecurity, CISASecureSoftwareAttestation, morrison2014mapping,sec_outcome,morrison2015security, zahan2023software}. While MTTR and MTTU are widely used industry metrics, our work is novel in applying them as outcome measures for security responsiveness. Faster updates reduce exposure windows, limiting attackers' opportunity to exploit disclosed weaknesses. We draw inspiration from the DevOps Research and Assessment (DORA) framework~\cite{dora_metrics}. DORA is one of the largest research studies in software engineering research, analyzing survey responses from over 31,000 professionals across six years (researched in the years 2014 through 2019)~\cite{forsgren2019accelerate}. DORA research identifies which software engineering practices, including security practices, improve software delivery performance. One of DORA's outcome metrics is a responsiveness metric that measures how long it takes a project to respond and deploy the corresponding updates. Research conducted in 2022 by the Google DORA team~\cite{google2022accelerate} specifically investigated SSC practices derived from the Secure Software Development Framework (SSDF)~\cite{souppaya2022secure} and Supply-chain Levels for Software Artifacts (SLSA)~\cite{SLSA}, and examined their impact on outcomes. The research has shown that SSC practices are positively associated with faster response and deployment times, and practitioners also anticipate a lower chance of security vulnerabilities. Motivated by DORA metrics, we hypothesize that \textit{teams that adopt good security practices are also more likely to be more responsive, as measured by MTTR and MTTU}. We used MTTU because ecosystem-wide MTTR data is often unavailable (only 15\% of our dataset; see section \ref{sec:sc-data-outcome}). While MTTR focuses on the response to security updates, MTTU measures responsiveness to any upstream updates and is available for all packages. Prior work~\cite{rahman2024characterizing} also found a moderate correlation between MTTR and MTTU, suggesting that MTTU can serve as a proxy when MTTR is unavailable.

\subsubsection{\textbf{Control Variables}} \label{control_vars}

Control variables are properties that researchers hold constant in an experiment to understand the true relationships between the predictor and target variables~\cite{FrostControlVariables}. We included six variables to measure repository characteristics based on two criteria: (1) alignment with prior use of control variables on software quality and practices work~\cite{miller2025understanding,zhou2019fork,ray2014large,ortu2019empirical,trockman2018adding,li2024can,javan2023dependency}, and (2) maintaining model interpretability with a minimal yet effective set of metrics to reduce data collection and cost. We controlled for six repository characteristic metrics grouped by their hypothesized influence. Contributor count (Contributors\_CT) and commit staleness reflect team capacity and development activity, which may affect the implementation and maintenance of security practices. Download count and direct dependencies capture project adoption and external exposure, potentially influencing both scrutiny and security risk. Repository age and size represent project maturity and complexity, which may impact maintenance effort and security performance. We excluded highly correlated variables such as forks, stars, and dependents due to redundancy and because they did not improve model performance.

\begin{table*}[t]
\renewcommand{\arraystretch}{1.2}
\captionsetup{skip=2pt}  
\caption{Scorecard Security Practices Metrics}
\vspace{-0.5ex}  
\caption*{\small
-1: no conclusive evidence or runtime error.  
0--10: extent of security practice adoption.  
Values in parentheses indicate raw counts.}
\vspace{-0.7ex}  
\label{tab:scorecards}
\centering
\begin{tabular}{|p{95pt} || p{260pt} ||p{30pt}|| p{30pt} || p{30pt} |}
\hline
Security Practices & Description & \textbf{-1 \%} & \textbf{0\%} & \textbf{1--10\%} \\
\hline\hline
Binary-Artifacts &Is the project free of checked-in binaries? & 0.0 & 0.2 & 99.8\\\hline
Branch-Protection & Does the project use Branch Protection?  & 0.0 (3) & 85.9 & 14.1\\\hline
Code-Review & Does the project practice code review before code is merged? & 0.4 & 75.4 & 24.2\\\hline
License & Does the project declare a license? & 0.0 & 30.9 & 69.1  \\\hline
Maintained & Is the project at least 90 days old and maintained? & 0.0 & 86.7 & 13.3\\\hline
Dependency-Update-Tool & Does the project use tools to help update its dependencies?  & 0.0 (65) & 81.3 & 18.7  \\\hline
Contributors & Does the project have contributors from different organizations?  & 0.0 & 36.7 & 63.3\\\hline
SAST & Does the project use static code analysis tools, e.g. CodeQL? & 0.0 (23) & 96.9 & 3.1 \\\hline
Security-Policy & Does the project contain a security policy? & 0.0 (2) & 93.2 & 6.8\\\hline
CI-Tests & Does the project run tests in CI, e.g. GitHub Actions, Prow? & 50.3 & 41.3 & 8.4\\\hline
Pinned-Dependencies & Does the project declare and pin dependencies? & 72.2 & 18.0 & 9.8\\\hline
Token-Permissions & Does the project declare GitHub workflow tokens as read-only?  & 73.2 & 25.6 & 1.2\\\hline
Dangerous-Workflow & Does the project avoid dangerous patterns in GitHub workflows?  & 73.2 & 0.3 & 26.6 \\\hline
Fuzzing & Does the project use fuzzing tools, e.g. OSS-Fuzz, QuickCheck? & 0.0 (2) & 99.8 & 0.2\\\hline
CII-Best-Practices & Has the project earned an OpenSSF Best Practices Badge? & 0.0 & 99.9 & 0.1\\\hline
Signed-Releases & Does the project cryptographically sign releases? & 97.7 & 2.3 & 0.0\\\hline
Packaging & Does the project build and publish official packages from CI/CD? & 97.5 & 0.0 & 2.5 \\\hline

Vulnerabilities & Does the project have unfixed vulnerabilities (including dependencies)?  & 0.0 & 32.2 & 67.8\\\hline

\end{tabular}
\end{table*}

\subsection{Dataset Source and Construction}\label{dataset-contruction}
Section \ref{sec:sc-data-scource} discusses different sources used to construct our dataset. Section~\ref{sec:exclusion-criteria}, Section~\ref {sec:sc-data-scorecard}, Section~\ref{sec:sc-data-outcome} and Section \ref{sec:sc-data-control} cover our dataset collection process. Our dataset is available in Figshare\footnote{\url{https://figshare.com/s/e74f94dfe1be6a90c4d1?file=54861686}}.

\subsubsection{\textbf{Data Sources}}\label{sec:sc-data-scource}
To collect data, we leveraged multiple sources.  We selected the npm ecosystem for this study because npm is the largest open-source ecosystem, comprising 37 million package versions, with 74\% of open-source codebases using JavaScript~\cite{ossra2024,sonatype2024}. Scorecard analyzes packages with source repositories hosted on GitHub~\cite{openssf_scorecard_2023} and periodically evaluates popular open-source projects. Within Scorecard's dataset of selected popular packages, 64\% are npm packages. Additionally, our security outcome metrics require analyzing project dependencies, specifically counting vulnerabilities in dependencies and measuring outdated and vulnerable dependency update patterns for MTTR and MTTU. To support our analysis, we collected the dependency graph of npm packages and mapped packages to their corresponding GitHub repositories using deps.dev~\cite{OSI}.  Deps.dev, developed and hosted by Google, provides comprehensive insights into OSS, including source code locations, dependency graphs, package metadata, licenses, releases, and vulnerabilities. Another platform used in our study is Open Source Vulnerabilities (OSV)~\cite{osv_adv}. The OSV platform aggregates vulnerabilities from various security advisories. Scorecard and Deps.dev depend on OSV to collect vulnerability information for projects and their dependencies. We also used the Libraries.io platform~\cite{LibrariesIO}, which collects publicly available OSS metadata scraped from the internet, including GitHub.

\subsubsection{\textbf{Package Inclusion and Exclusion Criteria}}\label{sec:exclusion-criteria}

To construct our dataset, we systematically filtered, cleaned, and processed npm packages based on dependency relationships, repository availability, and data availability. The dataset construction followed the following steps:

\begin{itemize}[leftmargin=10pt,itemindent=0pt]
    \item \textbf{Initial Dataset:} We began with a total of 2,119,044 unique npm packages extracted from the Deps.dev BigQuery dataset~\cite{depsdevBigQuery}. This dataset contains the mapping between npm packages and GitHub repositories and captures all dependency edges within npm.
    
    \item \textbf{Packages with GitHub Repositories:} Of these, 877,312 packages were excluded because they lacked valid GitHub repository information, either because they were hosted outside GitHub or had missing repository URLs. Mapping packages to GitHub repositories was necessary because Scorecard runs directly on GitHub repositories. After this filtering step, 1,241,732 packages remained that could be successfully mapped to GitHub.
    
    \item \textbf{Packages with Dependencies and Dependents:} Further filtering identified 1,159,891 packages with at least one declared dependency, ensuring relevance to our security outcome metrics. Additionally, we are primarily interested in packages that practitioners may use; hence, we refined the dataset further to include only packages with at least one dependent, reducing the dataset to 264,413 packages. 
    
    \item \textbf{Ensuring Unique Repositories:} Scorecard generates metric scores based on security practice adoption at the repository level rather than for individual packages. While we collected vulnerability data at the repository level (Section \ref{sec:sc-data-outcome}), MTTR and MTTU can only be measured at the package level. Consequently, if a repository contains multiple packages, all packages share the same Scorecard score and vulnerability data but may have different MTTR and MTTU values. Such inconsistency leads to duplicate rows with identical Scorecard metrics but differing security outcome metrics, which could mislead our machine learning model. To resolve this issue, we filtered the dataset to include only repositories that map to a single package. After applying this filtering step, we retained 172,999 unique repositories.

\end{itemize}

\subsubsection{\textbf{OpenSSF Scorecard Data}}\label{sec:sc-data-scorecard}
Once we identified the set of 172,999 unique npm GitHub repositories, we collected OpenSSF Scorecard scores for each repository associated with a package. Our primary data source was the OpenSSF Scorecard dataset in Google BigQuery~\cite{ScorecardBigQuery}. However, Scorecard only evaluates a subset of repositories, specifically popular packages and those explicitly requested by practitioners for inclusion. Hence, many repositories in our dataset were not included in the original Scorecard dataset. To address these gaps, we ran Scorecard locally for any repositories missing from the BigQuery dataset, using the same configuration applied to populate BigQuery. After merging data from Google BigQuery and our local runs, we obtained Scorecard scores for \totalpackagespace out of 172,999 repositories. The remaining repositories did not generate a score using Scorecard. Upon manual verification of a random sample of 100 repositories and their Scorecard output, we found that Scorecard failed to produce scores due to HTTP 404 errors. The primary reason for the lack of scores was that the repositories were no longer accessible via public GitHub URLs. The repositories were either private or had been removed from GitHub.

\textbf{Metrics exclusion}: We began by collecting all 18 metrics provided by Scorecard. However, due to limited adoption and Scorecard measuring limitations, we excluded Fuzzing, CII-Best-Practices, Signed-Releases, and Packaging from our analysis (Table~\ref{tab:scorecards}). Fuzzing and CII Best Practices were excluded because ~99.8\% of repositories scored 0, indicating minimal adoption of Scorecard recommendations across ecosystems. %
Signed Releases and Packaging were excluded, as almost 98\% of the repositories received a -1 score due to not having any signed release or a package release workflow in GitHub. Prior research~\cite{zahan2023openssf} found that practitioners primarily used package registries for releases rather than GitHub, and no improvements in Scorecard’s methodology addressing this limitation were observed. We also removed the vulnerability metric as a practice because we use \vulcountspace as a dependent variable. Thus, we removed five metrics from the analysis.

\subsubsection{\textbf{Security Outcome Metrics Data}} \label{sec:sc-data-outcome}
We collected our security outcome metrics for \totalpackagespace npm packages with Scorecard data. Here, we describe our outcome metrics data collection process.

\textbf{\vulcountspace metric}: The Scorecard Vulnerability metric measures whether the project has open, unfixed vulnerabilities in its codebase or its dependencies using the OSV~\cite{osv_adv} service. Initially, we used the Scorecard tool to collect the total vulnerability count. The Scorecard tool normalizes the vulnerability score, where a score of 10 indicates no vulnerabilities, 9 corresponds to one vulnerability, and 0 represents more than nine vulnerabilities. To obtain the exact vulnerability count, we first denormalized the score. For cases where the vulnerability score was 0 (indicating more than nine vulnerabilities), we ran the vulnerability metric separately using Scorecard to retrieve the raw, unnormalized data and used the total number of vulnerabilities in the model. 

\textbf{MTTR and MTTU metric}:
In this study, we adopt the algorithms proposed by Rahman et al.~\cite{rahman2024characterizing} for computing MTTR and MTTU. %
Following their methodology, we contacted the Googlee deps.dev team and considered deps.dev as our primary data source to collect comprehensive package versions,  dependency versions and their release information for our dataset. Additionally, we retrieved CVE data for selected packages and their direct dependencies from OSV~\cite{osv_adv}. Note that while MTTU captures all dependency updates, MTTR specifically accounts for updates addressing vulnerable dependencies. MTTR needs a vulnerability in direct dependency, then the availability of a new fix version, and the adoption time to adopt the new fix version by the dependent.  While we successfully collected \vulcountspace and MTTU data for \totalpackagespace packages, MTTR data was only available for \totalMTTRspace packages (15\% of total dataset). %

\subsubsection{\textbf{Control Variable Metrics Data}} \label{sec:sc-data-control}
We collected control variable data from different sources. GitHub contribution count and package size are collected from Libraries.io~\cite{LibrariesIO}. Then we got the number of downloads of a package in the past 12 months, which we collected from the public npm API\footnote{\url{https://api.npmjs.org/downloads/point/{period}[/{package}]}}. We calculated the repository age (in years) and commit staleness dates using the GitHub REST API~\cite{GitHubAPI2024}. The number of dependency information is collected from Deps.dev~\cite{depsdevBigQuery}.

\section{Data Preprocess} \label{sc:data-preporocess}

Before constructing the machine learning~(ML) modeling to answer RQs, we performed data preprocessing steps to handle missing values, ensure data integrity, and assess statistical properties. Section \ref{sc:data_imputation} covers feature decomposition and imputation techniques to manage missing values. Section \ref{sc:corelation} discusses correlation and multicollinearity in our dataset.

\subsection{\textbf{Feature Decomposition and Imputation}} \label{sc:data_imputation}

Our manual analysis found that Scorecard assigned -1 in two cases: (1) when there was no conclusive evidence of security practice adoption; and (2) due to internal runtime errors. Since not all metrics within a repository had a -1 score (Table \ref{tab:scorecards}), excluding an entire repository due to a -1 in any one of the 18 metrics would result in unnecessary data loss. To handle missing data effectively, we implemented a missing data pattern decomposition strategy based on Scorecard score reasoning. For high-missing features CI-Tests, Pinned-Dependencies, Token-Permissions, and Dangerous-Workflow (each with $>$50\% -1 scores), a score of -1 indicates that the repository lacks the relevant configuration. Here, -1 for CI-Tests indicates no pull requests to confirm CI workflows were present; for Pinned-Dependencies, it suggests the absence of declared dependencies in GitHub; for Token-Permissions, it reflects that no GitHub token settings were configured; and for Dangerous-Workflow, -1 means no GitHub Actions workflows were found to test workflow patterns. To preserve the distinction, we decomposed each feature into two components: a binary indicator of relevant configuration presence (1 for values 0-10, 0 for -1) and scores that capture adoption effectiveness (keeping original discrete values 1-10, with both -1 and 0 scores mapped to 0). For low-missing features ($<$1\% -1 values), where -1 was due to internal errors of Scorecard, we applied median imputation~\cite{schafer2002missing} to maintain data completeness while minimizing distortion.

\begin{figure*}[h]
    \centering
    \includegraphics[width=0.7\textwidth]{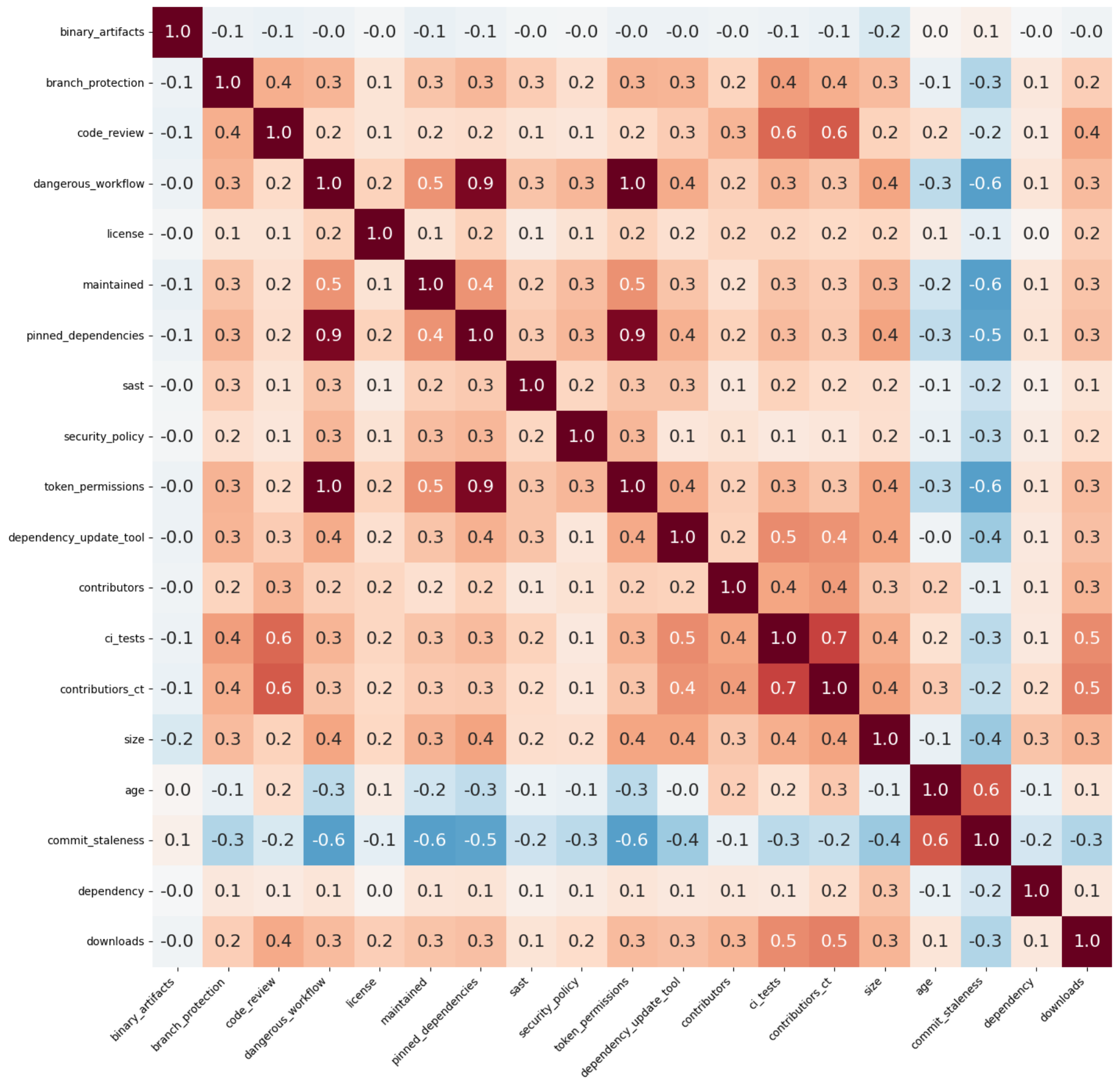}
    \caption{Spearman Correlation between Predictors}
    \label{fig:Correlation}
\end{figure*}

\subsection{\textbf{Correlation Analysis}} \label{sc:corelation}
To evaluate dataset normality, we applied the Henze-Zirkler multivariate normality test~\cite{henze1990class} separately to the RQ1 and RQ2 datasets. The RQ1 dataset includes aggregate scorecard scores, the control Variables, \vulcount, MTTR, and MTTU. The RQ2 dataset contains 13 Scorecard practice metrics, control variables, and \vulcount, MTTR, and MTTU. Both datasets failed the normality test (p $<$ 0.05), indicating deviations from a normal distribution. To further assess skewness, we flagged features with $|skewness| >$ 1 for potential transformation~\cite{hatem2022normality}. Based on skewness, we applied a log(x + 1) transformation to highly skewed control variables (GitHub contributors, repository size, DependencyCount, downloads) to mitigate skewness. Post-transformation, Q-Q plots, histograms, and statistical tests confirmed improved distribution symmetry. We did not apply transformation to the practice or outcome metrics to preserve their original scale and interoperability.  Since residual normality is important for regression analysis, we conducted D’Agostino-Pearson tests~\cite{keskin2006comparison} on regression residuals. Results showed that residuals remained non-normally distributed (p $<$ 0.05).

Given the non-normal distribution of our dataset, we proceeded with the Spearman correlation~\cite{zar2005spearman}, a nonparametric method suitable for capturing monotonic relationships in non-normal data. We conducted a correlation analysis for predictor variables (security practice and control variables) and target variables (security outcome metrics). Figure ~\ref{fig:Correlation} presents the Spearman correlation heatmap for predictor variables. The highest correlations (~1) were observed among Dangerous-Workflow, Pinned-Dependencies, and Token-Permissions. To avoid multicollinearity and redundancy in the model~\cite{guyon2003introduction}, we removed Dangerous-Workflow and Token-Permissions, retaining only Pinned-Dependencies as a representative feature. The correlation between target variables reveals a strong positive correlation (0.7) between MTTR and MTTU, aligning with our expectation of potential overlap~\cite{rahman2024characterizing}. In contrast, the \vulcountspace metric exhibits a weak negative correlation with MTTR (-0.3) and MTTU (-0.4). Hence, we confirmed that MTTR and MTTU cannot be used as target variables in the same model due to their strong correlation, since including both would introduce redundancy, bias, and multicollinearity. Additionally, we also computed the Variance Inflation Factor (VIF)~\cite{o2007caution} for all predictor variables to confirm multicollinearity. For the VIF computation, we did not find any feature that exhibited a VIF score greater than 5, indicating no multicollinearity. %

\section{RQ1:  Overall security posture and its security outcomes}
\label{sc:Rq1}
In this section, we discuss our \textbf{\rqPatternicse} Section \ref{rq1:method} discusses the methodology of RQ1, and Section \ref{rq1:result} discusses the results.

\begin{figure*}[htb] %
    \centering
    \includegraphics[width=\textwidth]{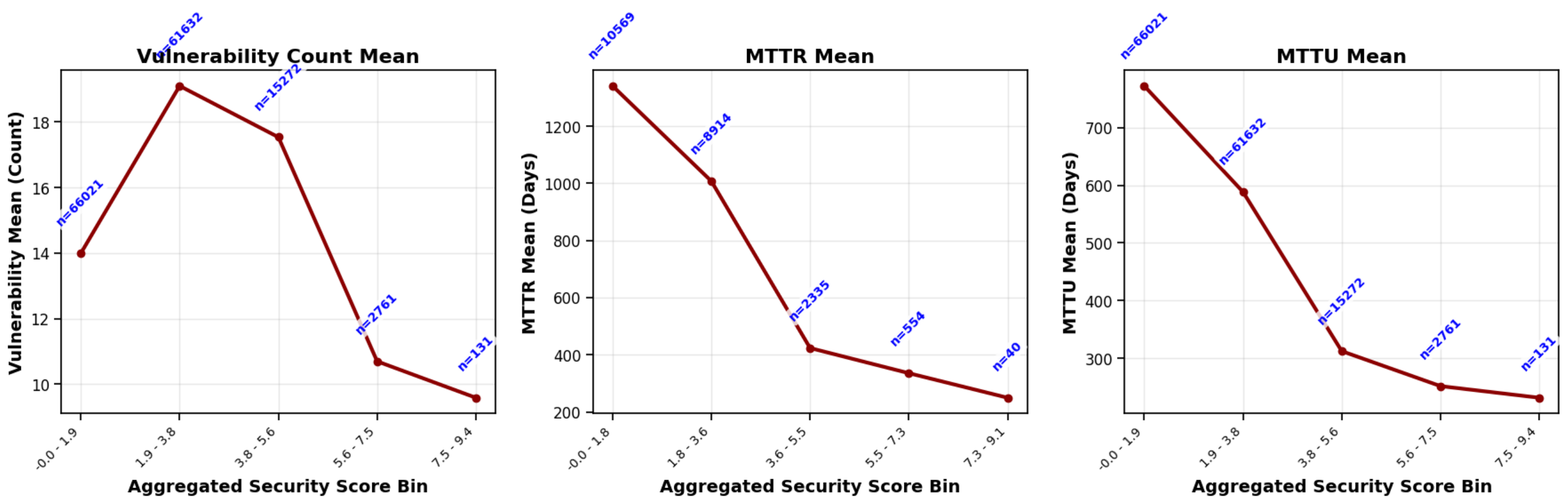}
    \caption{Security Outcome Trends and npm Repository Distribution Across Aggregated Scorecard Score Bins}
    \label{fig:security_outcomes_trend}
\end{figure*}
\subsection{RQ1: Methodology} \label{rq1:method}
For RQ1, we evaluate the relationship between the aggregated Scorecard score of npm package repositories and control variables as predictors, and security outcome metrics as the target variable. To visualize the relationship between aggregated Scorecard scores and security outcomes, we first binned repositories by aggregated Scorecard scores and analyzed mean security outcomes within each bin (Figure \ref{fig:security_outcomes_trend}). Next, we applied regression analysis to quantify the association of the aggregated Scorecard score with outcomes. However, regression analysis does not establish causality. We implemented a Propensity Score Matching (PSM) framework with caliper-based matching, covariate balance diagnostics, and segment-level analysis to estimate the causal effect of security practice adoption on outcome metrics, controlling for project characteristics. %

\subsubsection{\textbf{Data Expolation}} To visualize initial relationships, we binned repositories by aggregated Scorecard scores into five equal-width bins and calculated mean security outcomes within each bin. The binning approach provides an intuitive view of how security outcomes trend across different score ranges before applying statistical modeling.

\subsubsection{\textbf{Regression Model}:} 
At this stage, we focus on interpreting how security outcomes change per unit change in the aggregated Scorecard score (e.g., a coefficient of -0.1 implies a 10\% decrease in the outcome for every one-unit increase in aggregated Scorecard score).  We selected Generalized Linear Models (GLMs)~\cite{nelder1972generalized}, given the non-normal distribution of our dataset, and our target variables are count data. GLMs are explicitly designed for count data and provide interpretable coefficients. Coefficients quantify the expected change in the target variable for a unit increase in a predictor. To identify the best-fitting GLM, we systematically evaluated dataset characteristics using standard diagnostic tests: overdispersion (variance $<$ mean) and zero-inflation. We began with Poisson Regression~\cite{hausman1984econometric} as our baseline model and found significant overdispersion across all outcomes, with zero-inflation present only in \vulcount. We applied Negative Binomial (NB) Regression~\cite{hilbe2011negative} to address overdispersion and attempted Zero-Inflated Negative Binomial (ZINB)~\cite{ridout2001score} for \vulcountspace to handle both issues simultaneously. However, ZINB models encountered numerical convergence failures with precision loss warnings during maximum likelihood estimation. Therefore, we applied NB regression across all outcomes, providing stable parameter estimation while adequately addressing overdispersion.

\subsubsection{\textbf{High vs. Low Aggregated Scorecard Scores: Comparing Similar Repositories}} \label{rq1:causal_analysis}
To move beyond correlational evidence and establish causality, we employed PSM to simulate experimental conditions by comparing high and low-scoring repositories with similar project characteristics. PSM~\cite{austin2009balance} is a statistical technique that simulates the conditions of a randomized experiment by matching repositories with similar characteristics but differing in their level of security practice adoption. Similar project characteristics refer to repositories matched on the same covariates used as control variables in section \ref{control_vars}. 

Repositories are classified as ``treated" if their aggregated scores fall within the top quartile, with the remaining projects serving as a control group. We tested multiple treatment thresholds (median, tertile, quartile) and selected the top quartile ($\geq$ 3 on a 1-10 scale) to ensure sufficient contrast between treatment and control groups while maintaining adequate sample sizes. Using logistic regression on covariates that influence both aggregated scores and security outcomes, we estimated each repository's propensity score (probability of having a high aggregated score given its characteristics), then matched treated and control repositories with similar propensities. We then applied nearest-neighbor matching with a 0.1 standard deviation caliper to pair treated and control repositories with similar propensities. We validated matching quality using standardized mean differences (SMD) and calculated Average Treatment Effects (ATE) through difference-in-means with bootstrap confidence intervals (CI) to quantify uncertainty. We complement this analysis with NB on post-matched samples (containing only the matched treated and control repositories).
\begin{table}[htbp]
\centering
\caption{Segment-specific ATEs and percentage impacts. Strongest tercile per variable shown.}
\label{tab:RQ1_results_seg}
\begin{tabular}{|l@{\hskip 3.5pt}|l@{\hskip 3.5pt}|l@{\hskip 3.5pt}|l@{\hskip 3pt}|l@{\hskip 2pt}|}
\hline
\textbf{Outcome} & \textbf{Controls} & \textbf{Segment} & \textbf{ATE} & \textbf{Impact} \\
\hline
\textbf{Vul\_Count} & Contributors\_CT    & Many & -8.2   & 28.9\%       \\
                    & Age                 & Mature & -11.0   & 42.2\%       \\
                    & Commit Staleness    & Fresh & -9.4   & 47.5\%      \\
                    & Downloads           & High &  -9.3   & 35.9\%       \\
                    & Size                & Large & -8.4   & 27.7\%  \\
                    & Dependency          & Medium-Dep & -9.8   & 37.0\%      \\
\hline
\textbf{MTTR}       & Contributors\_CT    & Many & -350.1 & 41.7\%     \\
                    & Age                 & Mature & -429.8 & 37.5\%      \\
                    & Commit Staleness    & Fresh & -244.8 & 37.3\%       \\
                    & Downloads           & High & -269.5 & 40.5\%      \\
                    & Size                & Large & -315.0 & 40.2\%  \\
                    & Dependency          & Few & -245.7 & 28.1\%      \\
\hline
\textbf{MTTU}       & Contributors\_CT    & Many & -87.2 & 19.1\%       \\
                    & Age                 & Mature & -86.0  & 12.8\%      \\
                    & Size                & Small & 146.3  & 40.0\%  \\
                    & Commit Staleness    & Stale & 212.1  & 23.5\%       \\
                    & Downloads           & Few & 146.1  & 38.5\%       \\
                    & Dependency          & High-Dep & -81.0 & 20.6\%       \\
\hline
\end{tabular}
\end{table}
\subsubsection{\textbf{Impact Across Segments of Each Repository Characteristic}} While our PSM analysis establishes an overall causal effect by controlling for repository characteristics, it estimates one average treatment effect without examining how this effect varies across different repository types. For instance, the security benefits of higher aggregated scores might be stronger for large repositories compared to small ones.  To examine whether this effect differs in magnitude across different levels of repository characteristics, we conducted post-hoc heterogeneity analyses on the matched sample. We stratified the matched repositories into terciles based on each control variable using quantile-based binning and calculated segment-specific ATEs, comparing high and low scorecard scores within each segment. This approach identifies which repository contexts benefit most from higher aggregated scores. We used bootstrap CI to quantify uncertainty around segment-specific ATEs.

\subsection{RQ1: Result} \label{rq1:result}

\begin{table}[h]
\centering
\caption{Combined GLM and PSM analysis for RQ1 %
}
\label{tab:RQ1_results}
\begin{tabular}{|p{1.4cm}|p{1.2cm}|p{2.1cm}|p{2.6cm}|}
\hline
\textbf{Outcome} & \textbf{NB Coef} & \textbf{NB Coef (PSM)} &  \textbf{PSM ATE (95\% CI)}   \\
\hline
\textbf{\vulcount} & -0.20   & -0.26  & -5.2 [-5.7, -4.7]  \\
\textbf{MTTR}      & -0.05    & -0.16   & -216.8 [-241.8,-193.2] \\
\textbf{MTTU}      &  0.03   & -0.04   & -52.3 [-60.2, -44.8] \\
\hline
\end{tabular}
\end{table}
\textbf{Data Exploration}: Figure~\ref{fig:security_outcomes_trend} shows a sharp decline in security outcomes (indicating improvement) as aggregated Scorecard scores increase. The highest bin (7.3–9.1) reflects the top-scoring repositories, as no project achieved a perfect score. However, the plot only reflects an overall trend and does not account for confounding factors and associations.

\textbf{Quantitative Confirmation}: Table \ref{tab:RQ1_results} confirms visual trends (Figure~\ref{fig:security_outcomes_trend}) through three analytical approaches: full-sample regression, PSM post-matched sample regression, and PSM causal analysis. The columns show full-sample NB coefficients, post-matching NB coefficients, and post-matching ATEs with 95\% confidence intervals. For \vulcount, both full-sample (–0.20) and post-matching (–0.26) NB models show negative coefficients, with ATE indicating 5.2 fewer vulnerabilities. MTTR demonstrates consistent improvements: full-sample (–0.05), post-matching (–0.16), and ATE (–216.8 days). For MTTU, while the full-sample model shows a positive coefficient (0.03), post-matching (–0.04) and ATE (–52.3 days) reveal faster updates, suggesting confounding factors mask the relationship.

\textbf{Repository Context-Dependent Effects}: While our PSM analysis established overall causal effects, we conducted segment-specific analyses to understand where these effects are strongest. Table \ref{tab:RQ1_results_seg} presents the most pronounced effects within each repository characteristic, with the "Segment" column identifying the specific subgroup and the "Impact" column quantifying percentage improvements relative to the control baseline within that segment.

Across all segments, \textbf{\vulcountspace} consistently shows positive security outcomes (i.e., negative ATE values), with improvements ranging from 27.7\% to 47.5\%. The most substantial reductions in open vulnerabilities are observed in mature repositories with fresh commit staleness (47.5\% and 42.2\%, respectively), indicating that actively maintained and mature repositories with high aggregated scores tend to have fewer vulnerabilities. Medium-dependency group repositories show the greatest vulnerability reduction, highlighting the effect of dependency complexity.  \textbf{MTTR} improvements are also observed across all segments. Mature repositories have the largest absolute improvement (429.8 days faster), followed by those with many contributors (350.1 days). Repositories with fewer dependencies and fresh commit staleness show meaningful reductions as well (-245.7 and -244.8 days). Then, repositories with higher download counts tend to have fewer \vulcountspace and faster MTTR. \textbf{MTTU} reveals the most heterogeneous pattern, with some segments showing improvements while others exhibit deterioration. Small repositories by size and download and repositories with stale commit activity show increases in MTTU, suggesting that top quartile aggregated score repositories that are small, less downloaded, or less active have slower dependency updates compared to low Scorecard score repositories in the same segment category. Conversely, repositories with many contributors, mature age, and high dependency counts demonstrate faster updates.

\textbf{Summary of RQ1 Findings:}
\begin{itemize}[leftmargin=10pt,itemindent=0pt]
\item{\textbf{Overall Effects:}} Higher aggregated Scorecard scores improve security outcomes, with repositories achieving 5.2 fewer vulnerabilities, 216.8 days faster MTTR, and 52.3 days faster MTTU.
\item{\textbf{Repository Context Matters:}} The adoption of aggregated security practices improves security outcomes. The segment-specific analysis reveals nuanced patterns in how repository characteristics influence security practice effectiveness. 
\item{\textbf{Organizational Capacity:}} Mature repositories with many contributors consistently show better outcomes, suggesting that well-staffed and mature projects with high adoption of security practices benefit from improved security outcomes. 
\item {\textbf{Project size, activity, and downloads}}: Project size, recent commit activity, and download volume are associated with improvements in \vulcountspace and MTTR. However, their absence creates challenges (i.e., smaller, inactive, or less-downloaded projects) and correlates with slower MTTU, highlighting how inactivity and smaller projects with fewer downloads can hinder general updates. 
\item{\textbf{Dependency Complexity:}} Different dependency levels excel in different areas. Medium-dependency projects have the fewest open vulnerabilities, likely balancing manageable complexity. Few-dependency projects remediate issues fastest, benefiting from simpler, more transparent dependency trees. Large-dependency projects excel at general updates, likely leveraging automation to manage scale effectively.
\end{itemize}

\begin{tcolorbox}[colback=blue!5,colframe=blue!40!black] 
While selecting dependencies, organizations should prioritize mature, actively maintained large projects with many contributors, fewer dependencies, and high downloaded packages, as our analysis showed that such projects have fewer vulnerabilities and faster MTTR due to better security performance. Conversely, smaller, inactive packages with fewer downloads require enhanced monitoring and proactive security assessments to compensate for slower MTTU.
\end{tcolorbox}

\begin{table*}[htbp]
    \centering
    \caption{R$^2$, Security practice Importance Ranked and PSM Insights. Positive ATE indicates increased outcome (worse), negative ATE indicates reduced outcome (better).}
    \label{tab:feature_importance}
    \begin{tabular}{|p{40pt}|p{45pt}|p{205pt}|p{180pt}|}
    \hline
    \textbf{Outcome} & \textbf{R$^2$ for RF} & \textbf{Ranked Security Practices} & \textbf{PSM Effect (ATE direction)} \\ \hline
    \vulcount & 0.39 & Contributors, Code Review, License, Pinned-Dependencies,  Branch Protection, CI-Tests, Dependency-Update-Tool&  
    Code Review (+), CI-Tests~(mixed), Branch Protection (+), Dependency-Update-Tool (+), Others (-) \\ \hline
    MTTR & 0.50 & Contributors, Code Review, License, CI-Tests, Pinned-Dependencies &   Code Review (+), Others (-) \\ \hline

    MTTU & 0.32 & Contributors, License, Code Review, CI-Tests, Pinned-Dependencies &  
    Code Review(+), CI-Tests~(mixed), Others (-) \\ \hline
    \end{tabular}
\end{table*}

\section{RQ2: Impact of individual security practices and outcomes}\label{sc:Rq2}
In this section, we discussed our \textbf{\rqOutcomeicse} Section \ref{rq2:method} discusses the methodology for RQ2, and Section \ref{rq2:result} discusses the result. 

\subsection{RQ2: Methodology} \label{rq2:method}
For RQ2, we investigate the association between individual security practices and security outcomes. Unlike RQ1, where we analyzed aggregated Scorecard scores as a predictor, for RQ2, we analyzed eleven security practices as predictors after exclusion (Section \ref{sec:sc-data-scorecard} and \ref{sc:corelation}) alongside six control variables. Due to the large number of predictors, non-normal distribution, and non-linear relationships (Section \ref{sc:data-preporocess}), we selected a tree-based regression model suited for handling complex, high-dimensional data. We also conducted the feature importance ranked analysis to identify the security practices that most strongly contribute to predicting security outcomes. We performed causal analysis to explore potential causal relationships since regression and feature importance identify associations but do not establish causality. 

\subsubsection{\textbf{Regression Model:}} 
We employed Random Forest (RF) regression due to its robustness to non-linear relationships and ability to handle high-dimensional data without overfitting. Following preprocessing (section \ref{sc:data_imputation}), the dataset was split into training (70\%) and testing (30\%) subsets using random sampling. The model was trained on the training dataset using hyperparameters (n\_estimators=100, random\_state=42). To ensure robust performance estimation and reduce bias, five-fold cross-validation was applied exclusively to the training data, calculating R² scores(Table \ref{tab:feature_importance}).

\subsubsection{\textbf{Security Practice Ranking}}: 
Feature importance was quantified using Gini Importance, extracted directly from the trained Random Forest model's feature importances attribute, which measures each feature's contribution to reducing node impurity across all decision trees in the ensemble. Features were ranked by their grouped importance scores using dense ranking methodology (method= dense) within each dependent variable.  Since the preprocessing step decomposed high-missing features (CI-Tests, Pinned-Dependencies) into separate components, we grouped all feature importance scores by their base security practice to provide more interpretable ranking results. The method involved summing the decomposed components for CI-Tests and Pinned-Dependencies, while non-decomposed features retained their individual importance scores. Using the ranked importance scores of 11 practice metrics, we selected the top-performing security practices that collectively accounted for approximately 80\% of the cumulative feature importance across all three outcome metrics~\cite{ucla2023efa}. The approach provides interpretable insights into which security practices have the strongest overall predictive power for each outcome variable (Table \ref{tab:feature_importance}).

\subsubsection{\textbf{High vs. Low Security Practice Scores: Comparing Similar Repositories}} \label{rq2:causal_analysis}

We conducted PSM analysis on the top-ranked security practices from Table ~\ref{tab:feature_importance}. We focused on these top practices because the remaining ones demonstrated negligible importance values while the selected top features collectively accounted for approximately 80\% of the cumulative feature importance. However, individual security practices exhibit extreme binary distributions, with most repositories scoring either 0 (not adopted) or 10 (fully implemented), creating an extreme class imbalance that violates PSM overlap assumptions. Unlike aggregated Scorecard scores, which combine multiple practices to ensure nonzero values, individual practices remain at their adoption extremes. To enable valid causal inference, we applied balanced sampling to create sufficient propensity score overlap. Our PSM methodology employed a universal threshold of 5.0, classifying repositories with scores above this threshold as treated units and those at or below as control units. Following balanced sampling, after constructing balanced groups, we followed the technique discussed in Section \ref{rq1:causal_analysis}, where we used nearest-neighbor matching with a caliper adjustment and computed ATE. 

Unlike RQ1, where segmentation revealed how repository characteristics impact outcome, we could not perform a similar analysis in RQ2 due to data imbalance (Section~\ref{rq2:method}). The limited sample size in the PSM analysis further constrained our ability to isolate project-specific influences on security outcome metrics.

\subsection{RQ2: Result} \label{rq2:result}
\textbf{Quantitative Confirmation}: Table ~\ref{tab:feature_importance} summarizes our RQ2 findings, analyzing individual security practices' relationships with security outcomes. Security practices collectively explain a moderate portion of variance across outcomes: R² = 0.39 for \vulcountspace, R² = 0.50 for MTTR, and R² = 0.32 for MTTU. The higher explanatory power for MTTR suggests stronger relationships between security practices and MTTR compared to \vulcountspace or MTTU. %

\textbf{Important Security Practices:} The analysis reveals consistent patterns across outcome metrics. Contributors emerge as the most influential factor across all three outcome metrics, followed by Code Review and License practices. CI-Tests and Pinned-Dependencies consistently rank among the top practices. The ranking pattern suggests that collaborative human factors (Contributors) and development workflow practices (Code Review, CI-Tests) have an influential, measurable impact on security outcomes. Branch Protection and Dependency Update Tool ranks among the top seven features required to reach the 80\% cumulative importance threshold for \vulcount. In contrast, MTTR and MTTU reach the threshold with five features, but including these two additional practices raises the cumulative importance to nearly 92\% for MTTR and MTTU, indicating these practices also have a stronger impact on the outcomes. %

\textbf{Impact of Security Practice: } The PSM analysis reveals complex and counterintuitive relationships between individual security practices and outcomes. Code Review consistently exhibits a positive ATE across all three security outcomes, indicating associations with more vulnerabilities, slower MTTR, and MTTU. This paradoxical finding aligns with CISA's Secure by Design guidance~\cite{cisa2023securebydesign}, a rise in reported vulnerabilities can reflect a healthy security posture due to code review and testing by the community. CI-Tests show a mixed pattern: projects with pull-request (PR) infrastructure tend to have faster MTTR but also higher vulnerability counts and slower MTTU, possibly due to added complexity from collaboration. However, projects with PR infrastructure that also achieve high CI-Test scores consistently perform better across all security metrics. The finding suggests that PR workflows alone are insufficient; security benefits arise when effective CI practices are fully implemented.  Beyond Code Review and CI-Tests, other practices show more intuitive associations. Contributors, License, and Pinned Dependencies are associated with negative ATEs, suggesting improved outcomes if adopted by repositories. For \vulcount, the Branch Protection and Dependency Update Tool shows positive ATEs, suggesting these practices enhance vulnerability detection capabilities.

\begin{tcolorbox}[colback=blue!5,colframe=blue!40!black] 
Our analysis shows that Contributors, Code Review, License, CI-Tests, and Pinned-Dependencies are the most influential security practices across all outcome metrics. While adoption of Code Review practice shows positive associations with more vulnerabilities and slower MTTR and MTTU, Contributors, License, and Pinned Dependencies demonstrate negative ATEs, indicating improved security outcomes. CI-Tests show mixed effects depending on CI-test adoption. 
\end{tcolorbox}

\section{Discussion} \label{sc:disc}
The section discusses the research implications (section \ref{dis:takeaway}) and explores the threat to the validity (section \ref{dis:threat_to_validity}) of our study. 

\subsection {\textbf{Research Implication}}\label{dis:takeaway}
Defining and validating security outcome metrics is challenging; our work represents a step toward advancing that goal. Our study provides actionable insights for practitioners, policymakers, and researchers by investigating measurable security outcome metrics to assess the effectiveness of security practices.

\textbf{Holistic security practice adoption:}. RQ1 demonstrates that aggregated scores consistently improve security outcomes, while RQ2 reveals context-dependent and sometimes counterintuitive effects of individual practices. Our findings suggest that comprehensive security practice adoptions are more effective than isolated measures, as coordinated practices generate synergistic effects that individual actions alone cannot achieve. 

\textbf{Context-Aware Adoption:} The impact of security practice on security outcomes varies by project characteristics, requiring tailored strategies. While selecting dependencies, organizations should prioritize mature, actively maintained large projects with large contributors, fewer dependencies, and high downloaded packages, as our analysis showed that projects have fewer vulnerabilities and faster MTTR due to better security performance. Conversely, smaller or inactive packages with fewer downloads require enhanced monitoring and proactive security assessments to compensate for slower MTTU.

\textbf{Strategic Practice Prioritization:} Despite individual variability, Code Review, CI-Tests, License, Contributors, and Pinned Dependencies consistently demonstrate high feature importance across security outcomes. Our findings suggest that practitioners can adopt these empirically validated practices within their workflow, while recognizing that practices like Code Review, CI-test positive association reflect enhanced detection capabilities rather than security degradation. To ensure that security practice adoption is yielding tangible benefits, teams should regularly monitor outcome metrics. Continuous tracking of these indicators enables practitioners to assess the effectiveness of implemented practices and iteratively refine their security strategies based on empirical feedback.

\textbf{Evidence-Based Policy Development:} The quantified relationships between security practices and security outcomes provide empirical evidence for security frameworks and regulations. Policymakers can leverage these findings to develop guidelines that promote frameworks emphasizing coordinated, measurable practice adoption rather than checklist-based compliance. By aligning regulatory efforts with empirically supported metrics, policy can more effectively incentivize secure development behaviors across the software ecosystem.

\textbf{Future Work:} Our findings highlight the need to better understand the temporal feedback loop between security practices and outcomes. As projects adopt more practices, they may initially report more vulnerabilities and delays in dependency updates and fixing vulnerabilities due to improved detection rather than increased risk. The complexity calls for longitudinal studies that track how individual and aggregated practices affect security outcome metrics. For researchers, future work should model these dynamics in the context of project maturity, contributor activity, and development cycles to clarify whether observed security trends reflect degradation or deliberate, ongoing improvement. A comprehensive understanding of security outcome metrics is an open research challenge; therefore, researchers should explore new security outcome metrics beyond those used in this study.

\subsection{\textbf{Threat to validity}} \label{dis:threat_to_validity}
Our regression and causal analyses provide evidence supporting associations and causality between aggregated Scorecard scores, security practices, and security outcomes within our dataset. However, several inherent biases in our dataset must be acknowledged. First, our analyses are based on snapshot data capturing repository Scorecard metric scores and characteristics at a single point in time, lacking temporal granularity. Without longitudinal or time-series analysis, uncertainty remains about whether changes in security practices precede or directly influence observed security outcomes, thereby limiting the generalizability of our findings beyond our dataset. Second, selection bias may be present, as repositories adopting security practices could differ in unmeasured factors such as organizational policies, economic incentives, developer motivations, or project complexity. Although PSM mitigates confounding through observed covariates, unobserved variables may still influence both practices and outcomes.

Apart from that, our dataset exhibits other imbalances, as repositories have security practice scores of either 0, or -1, reflecting the limited adoption of these practices across the npm ecosystem. The real-world imbalance, characterized by the few repositories that adopt security practices, may introduce bias into our models, potentially skewing predictions or causal estimations. However, this imbalance highlights a widespread lack of security practice adoption among npm repositories. Hence, the imbalance reflects the actual state of adoption rather than a methodological flaw, highlighting the need for targeted interventions to increase the adoption of security practices among npm repositories. At the same time, the metric collection also needs improvement,  a team may be following a practice, but evidence of that practice may not be captured in the automated metrics. %

\section{Related Work} \label{sc:Related work}
This section highlights prior related studies, existing security guidelines, frameworks, and security outcome metrics.
 Prior studies~\cite{zahan2023software,sonatype_8} have utilized OpenSSF Scorecard metrics alongside vulnerability count data to understand the relationship between security practices and security outcome metrics. Zahan et al.~\cite{zahan2023software} analyzed 2,422 PyPI and npm packages and found a positive association between Scorecard metrics and vulnerability counts at the package level, but with low explanatory power ($R^2$ $<$ 12\%). They attributed this to limited dataset size, sparse vulnerability data at the package level, and missing control variables. Sonatype~\cite{sonatype_8} used a random forest classifier to predict vulnerable repositories based on Scorecard metrics and project attributes. However, binary classification limits understanding of the strength and direction of relationships between practices and outcomes. In contrast, our study employs regression and causal analysis to quantify associations, uses a larger, more representative dataset of \totalpackagespace npm packages, incorporates project-level outcome metrics, and controls for key project characteristics to improve validity and generalizability. 
 
\textbf{Security Practice Frameworks:} Executive Order (EO) 14028 emphasized adopting security practices to improve software security. To align with the EO, government agencies and industry experts have introduced various software supply chain security frameworks to enhance software supply chain security. These include the Software Component Verification Standard (SCVS) by OWASP~\cite{scvs}, the Building Security In Maturity Model (BSIMM)~\cite{BSIMM}, the CNCF Technical Advisory Group (TAG) recommendations~\cite{CNCF_TAG}, Supply Chain Levels for Software Artifacts (SLSA)~\cite{SLSA}, NIST Secure Software Development Framework (SSDF)~\cite{souppaya2022secure}, the Open Source Software (OSS) Secure Supply Chain (SSC) Framework by Microsoft~\cite{Microsoft_framework}, NIST’s Cybersecurity Supply Chain Risk Management Practices~\cite{NIST800-161}, the CISA Secure Software Development Attestation Form~\cite{CISASecureSoftwareAttestation} and OpenSSF Scorecard~\cite{Scorecard}. The Proactive Software Supply Chain Risk Management (P-SSCRM)~\cite{Williams2024PSSCRM} released in 2024, results from the study of real-world software supply chain risk management initiatives and the union of the 73 security practices in ten government and industry documents (standards and frameworks) mentioned above. In 2025, EO 14144~\cite{EO2025Cybersecurity} was issued, emphasizing the need for NIST to evaluate common cyber practices and security outcomes across industry sectors.

\textbf{Security Outcome Metrics:}Various metrics can assess the effectiveness of \practices by tracking security performance over time, generally categorized as threat-oriented, asset/impact-oriented, and vulnerability-oriented~\cite{NIST800-55,NIST800-30}. While threat- and asset/impact-oriented metrics offer valuable insights, they are difficult to quantify at scale due to reliance on real-time threat intelligence and complex impact assessments. In contrast, vulnerability-oriented metrics are more accessible and relevant to software security. Among them, severity-based metrics (e.g., CVSS) face challenges such as scoring inconsistencies and reliance on exploitability data~\cite{pendleton2016survey}. Temporal vulnerability metrics, such as vulnerability count, time to detect, time to repair and time to remediate, are widely used to evaluate security~\cite{morrison2014mapping,sec_outcome,CISA-2023,morrison2018mapping}. Morrison et al.~\cite{morrison2018mapping} conducted a systematic mapping study on software security metrics to evaluate the security properties of software and found that the most cited and most used metric is the vulnerability count (vulnerability metric). Davidson~\cite{davidson2009good} studied the time to close bug/vulnerability' (a form of mean time to repair) as a security metric and measured its effectiveness. Beres et al.~\cite{beres2009using} studied the window of exposure (vulnerability metric), which is the length of the interval of time needed by the security team to remediate a vulnerability once it has been disclosed. However, the scarcity of reported vulnerabilities, compounded by silently fixed issues, limits their reliability in assessing security outcomes~\cite{alhazmi2007measuring,dunlap2023finding}. To address these limitations, prior work suggests alternative metrics, such as fault prediction models~\cite{shin2013can}. Additionally, in software reliability, MTTR and MTTU have long been used to measure system resilience and maintenance efficiency~\cite{gokhale1999time,calvo2023applying}. While MTTR is often employed in security contexts~\cite{MTTR_1,MTTR_2,MTTR_3}, its reliance on reported vulnerabilities restricts its applicability. MTTU, by focusing on dependency updates, offers broader coverage but remains constrained by proprietary measurement methods~\cite{cogo2019empirical}.

\section{Conclusion} \label{sc:conclusion}
In this study, we investigated how the aggregate Scorecard score and Scorecard security practices metrics impact security outcome metrics. Our regression model and causal analysis reveal that higher aggregated Scorecard scores are associated with fewer vulnerabilities, shorter MTTR, and shorter MTTU. However, while regression models and causal analysis indicate better outcomes, our segment analysis to understand the impact of project characteristics suggests that factors beyond security practice adoption, such as project maintainability, resource availability, and dependency complexity, influence security outcomes. Feature ranking highlights Code Review, Contributors, License, CI-Tests, and Pinned Dependencies as top security practices associated with security outcomes, though their associations with security outcomes vary in direction. Additionally, MTTU exhibits trends similar to vulnerability count and MTTR, making it a potential proxy metric when vulnerability and MTTR data are unavailable. %

\bibliographystyle{acm}
\bibliography{arXiv}
\end{document}